\documentclass[pra,a4paper,nofootinbib,showpacs,aps,floatfix,superscriptaddress,twocolumn]{revtex4-2}
\usepackage[utf8]{inputenc}
\usepackage{graphicx}
\usepackage{bm}
\usepackage{amsmath}
\usepackage{color} 
\usepackage{amssymb}
\usepackage{bbold}
\usepackage{braket}
\usepackage{xcolor}

\DeclareMathOperator{\sinc}{sinc}

\graphicspath{{figs/}}

\begin{document}

\title{Exact Trotterization in the Degenerate Quantum Rabi Model}

\author{I. Lizuain}
\affiliation{Department of Applied Mathematics, University of the Basque Country UPV/EHU}
\affiliation{EHU Quantum Center, University of the Basque Country UPV/EHU}
\author{A. Rodriguez-Prieto}
\affiliation{Department of Applied Mathematics, University of the Basque Country UPV/EHU}
\affiliation{EHU Quantum Center, University of the Basque Country UPV/EHU}

\begin{abstract}
Digital quantum simulation of the quantum Rabi model in the deep strong coupling regime motivates the need for an exact characterization of Trotterization errors, so far known only through qualitative bounds. Here we derive such a characterization in the exactly solvable degenerate limit, valid for arbitrary coupling strength. The first-order Trotter operator admits a closed expression, showing that the generated phase-space trajectory lies exactly on a circle that is rescaled and rigidly rotated with respect to the exact one, rather than merely approximating it. This geometric renormalization yields a closed form for the state fidelity, structurally identical to the physical survival probability governing revivals in the deep strong coupling regime: both are the same geometric question — the distinguishability of a displaced Fock state — applied to two different pairs of points in phase space. We further show that this geometric error is exactly correctable by promoting the Trotter-step coupling to a complex amplitude, at no additional circuit cost. These results, obtained in the exactly solvable degenerate limit, provide a benchmark for Trotterization errors in the general Rabi model.
\end{abstract}

\maketitle

%
%

\section{Introduction}

The Quantum Rabi model (QRM) describes the interaction between a two-level system (a qubit) and a quantized bosonic mode, and lies at the heart of light-matter interaction theories \cite{Rabi1936}. While its rotating-wave approximation (RWA) — the Jaynes-Cummings model — captures the essential physics in the weak-coupling regime, it fails to describe the qualitatively different dynamics that emerge when the coupling strength becomes comparable to or exceeds the system frequencies. This Deep Strong Coupling (DSC) regime, once a theoretical construct, is now experimentally accessible in circuit QED and trapped-ion platforms~\cite{Casanova2010,Pedernales2015,Langford2017,Braumuller2017,FornDiaz2019,Lv2018}. There, the breakdown of the RWA gives rise to striking non-perturbative features with no counterpart in the Jaynes-Cummings model — most notably, collapse-and-revival dynamics even from Fock-like initial states, and highly non-classical qubit-oscillator states with large bosonic dressing.

Simulating this regime on a digital quantum processor requires decomposing the time-evolution operator into a sequence of elementary gates via a Trotter-Suzuki expansion~\cite{Lloyd1996,Ostmeyer2026}. However, the resulting Trotterization error is known to scale unfavorably with the coupling strength in the DSC regime, and its dependence on the simulation parameters has so far been characterized only through qualitative error bounds \cite{Childs2021}. An exact, closed-form characterization of this error — rather than a bound on its magnitude — remains an open problem.

Digital and digital-analog implementations of the QRM and related models have been explored on superconducting circuits and trapped ions~\cite{Mezzacapo2014,Pedernales2015}, and Trotterized simulations of the DSC regime have been studied numerically~\cite{Rochdi2024}, including recent benchmarks of low-order Trotter schemes for related light-matter models~\cite{Selvam2026}. To the best of our knowledge, none of these works provides a closed-form expression for the Trotter propagator itself, and consequently an exact analytical expression for the simulation fidelity has, so far, remained unavailable.

In this work we close this gap by exploiting the further simplification that occurs at $\omega_0=0$. While the full Rabi Hamiltonian is already known to be integrable for arbitrary $\omega_0$~\cite{Braak2011}, with its exact spectrum given implicitly by the zeros of a transcendental function, the degenerate limit $\omega_0=0$ considered here renders the model exactly diagonalizable in closed form by an elementary unitary transformation, valid for arbitrary coupling strength.
We derive a closed-form expression for the first-order Trotter operator, from which the resulting phase-space error is shown to have an exact geometric interpretation: the Trotterized trajectory does not merely approximate the exact one, but reproduces it exactly under a renormalized coupling and a fixed phase-space rotation. This renormalization yields, in turn, a closed-form expression for the state fidelity, which we show to share the same structure as the physical survival probability responsible for the revivals of Ref.~\cite{Casanova2010}: both reduce to the same geometric question — the distinguishability of a displaced Fock state — evaluated at two different pairs of points in phase space.
Although revival dynamics are most pronounced in the DSC regime, which motivates our focus, the underlying geometric mechanism is not tied to it. We further show that the identified error is exactly correctable by promoting the Trotter-step coupling to a complex amplitude, at no additional circuit cost. While our analysis is restricted to $\omega_0=0$, where these results are obtained in closed form, this limit captures the essential geometric mechanism at play and provides a benchmark against which Trotterization errors in the general Rabi model can be assessed.

\section{Theoretical Background}

The Deep Strong Coupling (DSC) regime of the quantum Rabi model is characterized by a set of distinctive non-perturbative features that fundamentally differ from those captured by the rotating wave approximation (RWA). In this regime, the interplay between the qubit and the bosonic mode gives rise to complex dynamics and highly non-classical states.
Among the most relevant phenomena is the emergence of revival dynamics from simple initial states. In the following, we briefly review this key aspect, originally presented in Ref.~\cite{Casanova2010} and further developed in Ref.~\cite{Pedernales2015}.

\subsection{The Quantum Rabi Model and the DSC regime}

The QRM describes the interaction between a two-level system
(a qubit) and a single bosonic mode. Its full Hamiltonian, without invoking the
rotating wave approximation (RWA), reads
\begin{equation}
    H = \omega a^\dagger a + \frac{\omega_0}{2}\sigma_z + g\,\sigma_x(a + a^\dagger),
    \label{eq:QRM}
\end{equation}
where $a^\dagger$ and $a$ are the bosonic creation and annihilation operators,
$\omega$ is the oscillator frequency, $\omega_0$ is the qubit energy splitting, and
$g$ is the coupling strength between the qubit and the field. The operators $\sigma_z$
and $\sigma_x$ are Pauli operators associated with the internal states of the qubit.
The interaction term $g\,\sigma_x(a + a^\dagger)$ includes both energy-conserving
(rotating) and energy-non-conserving (counter-rotating) contributions, distinguishing
the full Rabi model from its Jaynes-Cummings counterpart.

The DSC regime is defined~\cite{Casanova2010} by the condition $g \gtrsim \omega \gg \omega_0$. In this regime the RWA breaks down completely and the full Hamiltonian~\eqref{eq:QRM} must be considered. 
The DSC regime is usually characterized by a parameter $\lambda=g/\omega$ which describes the strength of the coupling with respect to the 
field frequency.

\subsection{The degenerate limit of the QRM}
\label{sec:diagonalization}
A particularly transparent limit of the Rabi Hamiltonian is $\omega_0 = 0$, in which the Hamiltonian reduces to

\begin{equation}
    H = \omega a^\dagger a + g\,\sigma_x(a + a^\dagger).
    \label{eq:QRM0}
\end{equation}
This form admits exact diagonalization via a qubit-controlled displacement unitary transformation. 
Let us define  the following unitary operator
$\mathcal{U} = \exp\!\left[\lambda\sigma_x(a^\dagger - a)\right]$ and compute interaction picture Hamiltonian $\widetilde{H} = \mathcal{U}H\mathcal{U}^\dagger$ using the BCH formula,
$e^G H e^{-G} = A + [G,A] + \frac{1}{2!}[G,[G,A]] + \cdots$, with $G = \lambda\sigma_x(a^\dagger - a)$.
The relevant commutators are
\begin{eqnarray}
\left[G,a\right]&=& \lambda \sigma_x[a^\dagger-a,\,a]= -\lambda\sigma_x,\\
\left[G,a^\dagger\right]&=& \lambda\sigma_x[a^\dagger-a,\,a^\dagger]= -\lambda\sigma_x,
\end{eqnarray}
and all higher-order commutators vanish identically, so the series terminates exactly at first order:
\begin{eqnarray}
\mathcal{U} a \mathcal{U}^\dagger &=& a- \lambda\sigma_x,\quad \mathcal{U} a^\dagger \mathcal{U}^\dagger = a^\dagger- \lambda\sigma_x.
\label{eq:at}
\end{eqnarray}
Substituting into $\widetilde{H} = \mathcal{U}H\mathcal{U}^\dagger$, the coupling
term cancel exactly and one obtains a diagonal, uncoupled Hamiltonian (up to a constant term not affecting the dynamics)
with the simple form $\widetilde H=\omega a^\dagger a$.

The unitary transformation $\mathcal{U}$  has a transparent 
physical interpretation: since $\sigma_x|\pm\rangle = \pm|\pm\rangle$, it acts 
as a qubit-controlled displacement operator $\mathcal{U}=D(\lambda\sigma_x)$ of the bosonic mode with a real parameter $\lambda=g/\omega$.
Applied to a separable state, it entangles the qubit with two coherent states of opposite 
amplitude, producing a superposition of the form $|+,\lambda\rangle + |-,-\lambda\rangle$, 
a Schrödinger cat-like state in phase space. This conditional displacement 
underlies the counterpropagating wavepacket picture of the DSC dynamics and the 
emergence of revival phenomena.

The full Rabi Hamiltonian is integrable for arbitrary $\omega_0$, in the sense of Braak's criterion based on its discrete parity symmetry $\Pi=-\sigma_z e^{i\pi a^\dagger a}$~\cite{Braak2011}, which for $\omega_0\neq0$ organizes the spectrum into two decoupled chains and underlies the emergence of revival dynamics from Fock-like initial states~\cite{Casanova2010}. The case $\omega_0=0$ considered here — the degenerate limit, since the qubit states $\ket{g}$ and $\ket{e}$ become degenerate — is special not because it restores integrability, already guaranteed by Braak's result, but because it renders the model exactly diagonalizable in closed form by the elementary unitary transformation $\mathcal{U}$ above, with no transcendental equation to solve. For $\omega_0\neq0$, $\mathcal{U}$ no longer diagonalizes the Hamiltonian, since $[\mathcal{U},\sigma_z]\neq0$, and the BCH series for $\widetilde{H}$ no longer truncates at first order.

For $\omega_0\neq0$ this displacement mechanism and the associated revival dynamics persist only approximately, as partial revivals with slowly drifting phases~\cite{Casanova2010}, reflecting the loss of exact solvability in closed form. In the rest of this work we restrict ourselves to the $\omega_0=0$ limiting case, which admits the exact analytical treatment developed below and allows us to isolate, in closed form, the essential geometric mechanism underlying both the physical revivals and the Trotterization error.

\subsection{Quantum revivals from Fock states}
\label{sec:revivals}

A direct consequence of the displaced-oscillator structure is the 
emergence of revival dynamics for initial states that are not eigenstates 
of the full Hamiltonian~\cite{Casanova2010}. In the limit $\omega_0 = 0$, 
the survival probability can be computed exactly.

Let us consider an initial pure Fock state with $n$ photons in the form 
$\ket{\psi_0}=c_+\ket{+,n}+c_-\ket{-,n}$, normalized by 
$|c_+|^2+|c_-|^2=1$, and compute the survival probability as the 
projection of the time-evolved state onto the initial state,
\begin{equation}
 \mathcal{S}(t) = \left|\bra{\psi_0}e^{-iHt}\ket{\psi_0}\right|^2.
 \label{eq:survival_prob_general}
\end{equation}
As shown in Appendix~\ref{app:survival}, this probability reduces to a 
single bosonic matrix element,
\begin{equation}
 \mathcal{S}(t) = \left|\bra{n}D[\alpha(t)]\ket{n}\right|^2,
 \qquad \alpha(t) \equiv \lambda\left(e^{-i\omega t}-1\right).
\end{equation}
As expected from the degeneracy of $\ket{g}$ and $\ket{e}$ at $\omega_0=0$, $\mathcal{S}(t)$ is independent of the qubit state, i.e., of $c_+$ and $c_-$: the two qubit eigenstates $\ket{\pm}$ contribute identically to the projection, regardless of the initial qubit superposition.

\begin{figure}[t]
\centering
\includegraphics[width=\columnwidth]{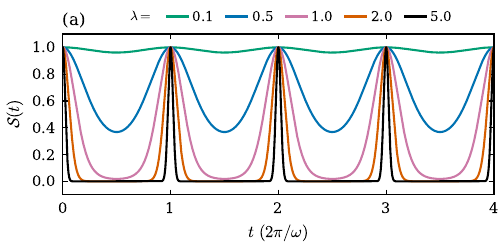}
\includegraphics[width=\columnwidth]{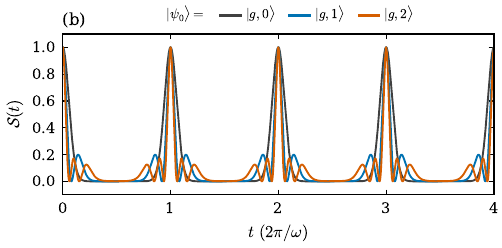}
\caption{Revival probability $\mathcal{S}(t)$ as a function of time in the degenerate limit $\omega_0=0$.
(a) Initial state $\ket{g,0}$ for different coupling strengths $\lambda=g/\omega$, ranging from the SC to the DSC regime of the QRM.
(b) Fixed coupling strength $\lambda=2$ for increasing photon numbers $n$ in the initial state $\ket{g,n}$, showing secondary peaks separated by the zeros of the Laguerre polynomial $L_n$.
In both panels, $\mathcal{S}(t)$ is independent of the initial qubit state: any superposition $c_+\ket{+,n}+c_-\ket{-,n}$ yields the same dynamics, as shown in the text.}
\label{Pg0_revivals_fig_label}
\end{figure}

Using the standard diagonal matrix element of the displacement operator in the Fock basis~\cite{WallsMilburn}, the survival probability takes the closed form
\begin{equation}
 \mathcal{S}(t) = e^{-|\alpha|^2}\left[L_n(|\alpha|^2)\right]^2,
 \label{eq:survival_Ln}
\end{equation}
with $L_n(x)$ being the Laguerre polynomial of order $n$. The survival 
probability is therefore entirely determined by $|\alpha|^2$, the
squared phase-space distance between the bosonic amplitude at time $t$ 
and its initial value at $t=0$,
\begin{equation}
 |\alpha|^2 = 4\lambda^2\sin^2\frac{\omega t}{2}.
 \label{eq:alpha_square}
\end{equation}
This quantity vanishes whenever $t=2\pi k/\omega$ ($k\in\mathbb{Z}$), 
producing the main periodic revivals. For $n=0$, $L_0=1$ and the pattern 
is purely Gaussian; for $n>0$, the zeros of $L_n(|\alpha|^2)$ generate
secondary peaks, reflecting interference between counterpropagating 
wavepackets~\cite{Casanova2010}, see Fig. \ref{Pg0_revivals_fig_label}.

\section{Exact Trotterization in the degenerate limit}
\label{sec:trotter}

Digital quantum simulation of the QRM proceeds by approximating the time-evolution operator $e^{-iHt}$ by a sequence of simpler unitaries.
The key idea is to decompose $H$ into a sum of terms and approximate the joint evolution by a product of individually implementable 
exponentials --- the Trotter-Suzuki decomposition. In the degenerate limit $\omega_0=0$ considered here, the Hamiltonian reduces to 
$H = \omega a^\dagger a + g\,\sigma_x(a+a^\dagger)$, and the first-order Trotter step at time $\delta t$ takes the two-factor form
\begin{equation}
    e^{-iH \delta t} \approx \mathcal{T} =
    e^{-i\omega a^\dagger a\,\delta t}\;
    e^{-ig\sigma_x(a+a^\dagger)\delta t},
    \label{eq:trotter_step}
\end{equation}
repeated $k$ times to cover a simulation time $t_k = k\delta t$,
\begin{equation}
    e^{-iHt_k} \approx \mathcal{T}^k.
\end{equation}
The approximation error arises from the non-commutativity of the two factors, and is quantified by the state fidelity between the exact and 
Trotterized evolutions.
As we show in the following, in this degenerate limit $\mathcal{T}^k$ admits a closed-form expression that allows for an exact analytical 
computation of the fidelity, going well beyond qualitative error bounds.


The Trotter-step operator in Eq.~(\ref{eq:trotter_step}) consists of two elementary transformations. 
The oscillator term corresponds to a rotation of the bosonic mode in phase space by an angle $\theta\equiv\omega \delta t$,
while the interaction term can be rewritten as
\begin{equation}
    e^{-ig\sigma_x(a+a^\dagger)\delta t}=e^{\sigma_x(\beta a^\dagger-\beta^* a)}=D(\beta\sigma_x)
\end{equation}
and can therefore be identified as a qubit-conditioned displacement operator with imaginary amplitude 
$\beta\equiv-ig\delta t=-i\lambda\theta$.
The first-order Trotter step can therefore be written as
\begin{eqnarray}
 \mathcal{T}=e^{-i\theta a^\dagger a} D(\beta\sigma_x).
 \label{eq:trotter_step1}
\end{eqnarray}
Each Trotter step thus consists of a qubit-controlled displacement followed by a phase-space rotation of the bosonic mode. Repeated application of $\mathcal{T}$ generates a discrete trajectory in phase space, where the oscillator state is successively displaced and rotated at every step. In the continuum limit $\delta t\to0$, these discrete transformations converge to the exact continuous evolution, while for finite $\delta t$ they generate a polygonal approximation whose deviation from the exact trajectory gives rise to the Trotter error.

Using now the standard relations
\begin{equation}
 e^{i\theta a^\dagger a}a e^{-i\theta a^\dagger a}=a e^{-i\theta},\quad
 e^{i\theta a^\dagger a}a^\dagger e^{-i\theta a^\dagger a}=a^\dagger e^{i\theta},\nonumber
\end{equation}
one finds how the displacement operator transforms under a phase-space rotation,
\begin{equation}
 e^{i\theta a^\dagger a}D(\beta\sigma_x)\,e^{-i\theta a^\dagger a}=D(\beta e^{i\theta}\sigma_x),
 \label{eq:rot_disp}
\end{equation}
i.e., it simply rotates the displacement amplitude by an angle $\theta$ in phase space. Equivalently,
\begin{equation}
 e^{-i\theta a^\dagger a}\, D(\beta\sigma_x) = D(\beta e^{-i\theta}\sigma_x)\, e^{-i\theta a^\dagger a}.
 \label{eq:rot_disp_reversed}
\end{equation}
In other words, the Trotter step $\mathcal{T}$ in Eq.~(\ref{eq:trotter_step}) can be written indistinctly with the rotation on the left or on the right of the displacement, at the price of rotating the displacement amplitude accordingly.

Using this relation, it is instructive to build up $\mathcal{T}^k$ explicitly for the first few steps and identify the pattern by inspection. 
The second Trotter step is given by
\begin{eqnarray}
    \mathcal{T}^2 &=& D(\beta e^{-i\theta}\sigma_x)\, e^{-i\theta a^\dagger a}\, D(\beta e^{-i\theta}\sigma_x)\, e^{-i\theta a^\dagger a}\nonumber\\
    &=& D(\beta e^{-i\theta}\sigma_x)\, D(\beta e^{-2i\theta}\sigma_x)\, e^{-2i\theta a^\dagger a}\nonumber\\
    &=& D\!\left[\beta\!\left(e^{-i\theta}+e^{-2i\theta}\right)\sigma_x\right] e^{-2i\theta a^\dagger a},
\end{eqnarray}
where in the last line we composed the two displacements.  
Repeating the same manipulation, and again pushing the accumulated rotation $e^{-2i\theta a^\dagger a}$ through the next displacement,
\begin{equation}
    \mathcal{T}^3 = D\!\left[\beta\!\left(e^{-i\theta}+e^{-2i\theta}+e^{-3i\theta}\right)\sigma_x\right] e^{-3i\theta a^\dagger a}.
\end{equation}
Each step adds one further rotated copy of $\beta$ to the displacement amplitude while advancing the accumulated rotation by one unit of $\theta$.  After $k$ steps one thus obtains, by direct extrapolation of the pattern above,
\begin{eqnarray}
    \mathcal{T}^k &=& D(\gamma_k\sigma_x)\, e^{-i k \theta a^\dagger a}, \quad
    \gamma_k = \beta\sum_{j=1}^k e^{-i j \theta},
    \label{eq:Tk_exact}
\end{eqnarray}
i.e., the net displacement amplitude is the coherent sum of all individual Trotter-step contributions, each rotated by an additional angle $\theta$ relative to the previous one.

The geometric series in the equation above is readily summed to give
\begin{equation}
    \gamma_k = \beta e^{-i\theta}\frac{1-e^{-ik\theta}}{1-e^{-i\theta}},
    \label{eq:gamma_sum}
\end{equation}
and this expression can be brought into a more transparent form by recalling $\beta=-i\lambda\theta$ and the exact phase-space amplitude $\alpha(t)=\lambda(e^{-i\omega t}-1)$, which gives
\begin{equation}
\label{eq:gammak}
\gamma_k= \frac{e^{-i\theta/2}}{\sinc(\theta/2)}\,\alpha(t_k).
\end{equation}
The full $k$-step Trotter evolution at time $t_k=k\delta t$ thus consists of a rigid phase-space rotation together with a net qubit-conditioned displacement $\gamma_k$, tracing a polygonal trajectory in the complex plane as $k$ increases. This amplitude is directly proportional to the exact one, $\alpha(t_k)$, with a $k$-independent factor depending only on $\theta$, Eq.~(\ref{eq:gammak}). In the continuum limit $\theta\to0$ this factor tends to one and the exact dynamics is recovered, $\gamma_k\to\alpha(t_k)$. For finite $\theta$, this proportionality has a direct geometric consequence: instead of merely approximating the exact circular trajectory, the polygon vertices $\gamma_k$ lie exactly on a second circle, rescaled and rotated with respect to the exact one, as examined in the following sections.

\section{Geometric structure of the Trotter error}

Having established the closed-form expression for $\mathcal{T}^k$, we now turn to the geometric origin of the Trotterization error. To this end, we analyze the phase-space trajectories generated by the expectation value of the bosonic operator $\langle a(t)\rangle$ under both the exact and Trotterized dynamics. Rather than evaluating this quantity for specific initial states, it is more illuminating to first derive the exact and Trotterized Heisenberg evolution of the operator $a$ itself, from which the error and its geometric structure follow directly.

\subsection{Exact Heisenberg evolution of $a$}

The Heisenberg evolution of $a$ under the exact dynamics is given by $a(t)_\mathrm{ex} = e^{iHt}\,a\,e^{-iHt}$.
Using the diagonalizing transformation $\mathcal{U}=D(\lambda\sigma_x)$, which satisfies $\mathcal{U}H\mathcal{U}^\dagger=\widetilde{H}=\omega a^\dagger a$ and $\mathcal{U}a\mathcal{U}^\dagger = a-\lambda\sigma_x$
(see Sec.~\ref{sec:diagonalization}), this can be evaluated exactly as follows:

\begin{eqnarray}
 a(t)_{\textrm{ex}}&=&e^{iHt}ae^{-iHt}=\mathcal{U}^\dagger e^{i\widetilde Ht}(a-\lambda \sigma_x) e^{-i\widetilde Ht} \mathcal{U}\nonumber\\
 &=&\mathcal{U}^\dagger e^{i\widetilde Ht}a e^{-i\widetilde Ht} \mathcal{U}
 - \lambda\mathcal{U}^\dagger e^{i\widetilde Ht} \sigma_x e^{-i\widetilde Ht} \mathcal{U}\nonumber\\
 &=&e^{-i\omega t} \mathcal{U}^\dagger a \mathcal{U}-\lambda \sigma_x \nonumber\\
 &=&e^{-i\omega t} (a+\lambda \sigma_x)-\lambda \sigma_x\nonumber\\
 &=&ae^{-i\omega t}+\alpha(t)\sigma_x
 \label{eq:a_exact}
\end{eqnarray}
The exact Heisenberg operator $a(t)_\mathrm{ex}$ thus splits naturally into two contributions: a free oscillator term $ae^{-i\omega t}$, 
rotating at frequency $\omega$,
and a term $\alpha(t)\sigma_x$ representing a qubit-conditioned displacement of amplitude $\alpha(t)$, which couples the bosonic and qubit degrees of freedom through $\sigma_x$.
The structure of this result reflects directly the diagonalization process: the free rotation comes from $\widetilde{H}$, while the displacement 
term is the fingerprint of the transformation $\mathcal{U}$ that connects the original and diagonal frames. Notably, $\alpha(t)$ emerges here once again as the natural phase-space amplitude of the problem, the same quantity that governed the survival probability in Sec.~\ref{sec:revivals} and the closed-form Trotter operator in Sec.~\ref{sec:trotter} — a recurrence that anticipates the direct comparison between exact and Trotterized dynamics developed below.

\subsection{Trotterized evolution of $a$}

The Heisenberg evolution of $a$ under the Trotterized dynamics after $k$ steps is given by
\begin{equation}
 (a_k)_\mathrm{tr} = (\mathcal{T}^\dagger)^k\, a\, \mathcal{T}^k.\nonumber
\end{equation}
Using the closed-form expression $\mathcal{T}^k$ in Eq. (\ref{eq:Tk_exact})
\begin{eqnarray}
 (a_k)_\mathrm{tr} &=& e^{ik\theta a^\dagger a} D^\dagger(\gamma_k\sigma_x) a  D(\gamma_k\sigma_x) e^{-ik\theta a^\dagger a}\nonumber\\
 &=&e^{ik\theta a^\dagger a}(a+\gamma_k \sigma_x)e^{-ik\theta a^\dagger a}\nonumber\\
 &=&a e^{-ik\theta}+\gamma_k\sigma_x,
  \label{eq:a_trotter_heisenberg}
\end{eqnarray}
where we have used the standard displacement identity $D^\dagger(\xi)\,a\,D(\xi) = a + \xi$~\cite{WallsMilburn}.

This result has exactly the same structure as the exact operator, Eq.~(\ref{eq:a_exact}), with the continuous rotation $e^{-i\omega t}$ replaced by its discrete counterpart $e^{-ik\theta}$ with $\theta=\omega \delta t$, and the exact displacement amplitude $\alpha(t)$ replaced by the Trotterized amplitude $\gamma_k$. This one-to-one correspondence is a direct consequence of $\mathcal{T}^k$ sharing the same rotation-plus-displacement structure as the exact diagonalized evolution, and is precisely what allows the Trotterization error to be captured by a single complex number, as shown next.

\subsection{Phase-space error and geometric renormalization}

\subsubsection{Geometric interpretation of the error amplitude $\Delta_k$}

\begin{figure}[t]
\centering
\includegraphics[width=\columnwidth]{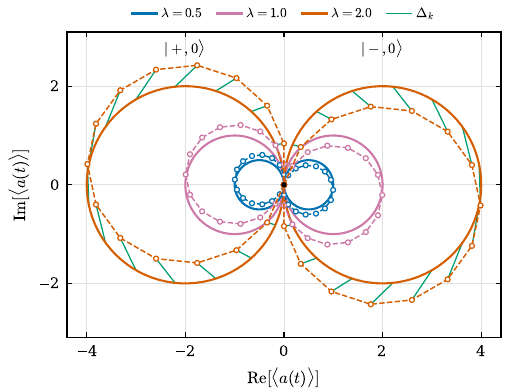}
\caption{Phase-space trajectories of $\langle a(t)\rangle$ over one full period $T = 2\pi/\omega$ for the exact evolution (thick continuous curves) and the first-order Trotterized dynamics with $15$ Trotter steps (dashed polygonal paths with open circles at the vertices $\gamma_k$), shown for increasing coupling strengths $\lambda = 0.5,\, 1.0,\, 2.0$ (blue, pink, orange).
The calculations are performed for initial states $\ket{+,0}$ and $\ket{-,0}$, which have maximal qubit polarization $\langle\sigma_x\rangle=\pm1$ and therefore trace orbits of maximum radius $\lambda$.
All trajectories start and end at the origin (black dot), since $\langle a\rangle=0$ at $t=0$ and both the exact and the Trotterized amplitudes return to zero after one full period.
Trajectories displaced to the left (right) correspond to $\ket{+,0}$ ($\ket{-,0}$). For a general initial state, the radius is modulated by the polarization $|c_+|^2-|c_-|^2$, and the trajectories are identical for any Fock number $n$.
The green segments shown for $\lambda=2.0$ connect each Trotterized point $\gamma_k$ to the corresponding point $\alpha(t_k)$ on the exact orbit at the same time, illustrating the phase-space separation $\Delta_k$ defined in Eq.~(\ref{eq:Deltak}).}
\label{fig:phase_space_a}
\end{figure}

The exact and Trotterized Heisenberg evolutions derived above share the same operator structure: both consist of a free bosonic rotation together with a qubit-conditioned displacement in phase space. Their difference thus defines a natural error operator,
\begin{equation}
 \delta a(t_k) = a(t_k)_\mathrm{ex} - (a_k)_\mathrm{tr} = \Delta_k\,\sigma_x,
 \label{eq:Deltak}
\end{equation}
where $\Delta_k \equiv \gamma_k-\alpha(t_k)$. The Trotterization error is therefore entirely encoded in the single complex number $\Delta_k$, which measures the complex phase-space distance between the exact and Trotterized displacement amplitudes at time $t_k$. Since $\delta a(t_k)$ contains neither $a$ nor $a^\dagger$, the error is entirely independent of the initial photon number, and enters observables only through the qubit polarization $\langle\sigma_x\rangle$, which is a conserved quantity as $[\sigma_x,H]=0$ at $\omega_0=0$.

To illustrate the geometric meaning of $\Delta_k$, let us evaluate the expectation values of the bosonic operator for an initial state $\ket{\psi_0}=c_+\ket{+,n}+c_-\ket{-,n}$. Using Eqs.~(\ref{eq:a_exact}) and (\ref{eq:a_trotter_heisenberg}), and $\langle a\rangle=0$ for any Fock state,
\begin{eqnarray}
 \langle a(t)\rangle_\mathrm{ex} &=& \big(|c_+|^2-|c_-|^2\big)\alpha(t),\\
 \langle a_k\rangle_\mathrm{tr} &=& \big(|c_+|^2-|c_-|^2\big)\gamma_k.
\end{eqnarray}
Both are governed by the conserved qubit polarization $\langle\sigma_x\rangle=|c_+|^2-|c_-|^2$: as shown in Fig.~\ref{fig:phase_space_a}, the exact evolution traces a circle of radius $r=\lambda\langle\sigma_x\rangle$ centered at $(-r,0)$ — vanishing for $\ket{g}$ or $\ket{e}$ and maximal, $r=\lambda$, for the eigenstates $\ket{\pm}$ — while the Trotterized evolution produces a polygon whose vertices converge to this circle as $\delta t\to0$. This polygonal structure follows directly from the Trotter decomposition: each step contributes a displacement rotated by a further angle $\theta$ relative to the previous one, so that $\gamma_k$ traces a discrete approximation to the continuous orbit.

Geometrically, $\Delta_k$ is the vector connecting each polygon vertex $\gamma_k$ to the corresponding point $\alpha(t_k)$ on the exact circle at the same instant, giving the Trotterization error a direct interpretation as a phase-space distance between exact and simulated dynamics. This distance vanishes in the continuum limit, grows with the coupling strength $\lambda$, and returns close to zero whenever $t_k$ approaches a multiple of $2\pi/\omega$, when both trajectories simultaneously complete a full period.

\subsubsection{Renormalized dynamics}
\label{sec:effective}

\begin{figure}[t]
\centering
\includegraphics[width=\columnwidth]{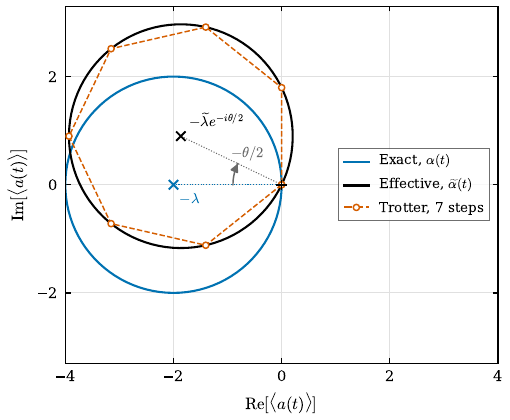}
\caption{Phase-space trajectories of $\langle a(t)\rangle$ for the
exact evolution (blue circle), the first-order Trotterized simulation
with 7 steps (red polygon), and the continuously evolved
renormalized model with coupling $\widetilde\lambda$ and phase
$-\theta/2$ (black circle). The polygon vertices fall exactly on the
black circle, confirming that the Trotterized dynamics reproduces
the renormalized model exactly. Crosses mark the exact and effective
circle centers, $-\lambda$ and $-\widetilde\lambda e^{-i\theta/2}$,
the latter rotated by $-\theta/2$ relative to the former, as indicated
by the arc. Parameters: $\lambda=2$, initial state $\ket{+,0}$.}
\label{fig:effective_circle}
\end{figure}

The closed-form expression for $\gamma_k$ in Eq.~(\ref{eq:gammak})
reveals a deeper structure of the Trotterized dynamics. Dividing
$\gamma_k$ by the exact amplitude $\alpha(t_k)$, one finds
\begin{equation}
 \frac{\gamma_k}{\alpha(t_k)} = \frac{e^{-i\theta/2}}{\sinc(\theta/2)}
 \equiv \varepsilon(\theta),
 \label{eq:ratio}
\end{equation}
independent of $k$: the Trotterized amplitude $\gamma_k$ is not merely
an approximation of the exact amplitude $\alpha(t_k)$, it is exactly
proportional to it at every time step, with a complex factor
$\varepsilon(\theta)$ depending only on the Trotter step size $\theta=\omega \delta t$.
The phase-space separation $\Delta_k$, Eq.~(\ref{eq:Deltak}), then takes
the compact form
\begin{equation}
 \Delta_k = [\varepsilon(\theta)-1]\,\alpha(t_k),
 \label{eq:Delta_factored}
\end{equation}
directly proportional to $\alpha(t_k)$ and vanishing in the continuum
limit $\theta\to0$ — consistent with the error being largest when the
system is far from the origin in phase space, and vanishing automatically
at revival times.

The geometric picture is particularly transparent: instead of merely
approximating the exact circular trajectory, the vertices $\gamma_k$
lie exactly on a second circle
$\widetilde{\alpha}(t) = \varepsilon(\theta)\,\alpha(t)$. This
effective circle has radius
\begin{equation}
 \widetilde{\lambda} = \frac{\lambda}{\sinc(\theta/2)} > \lambda,
 \label{eq:lambda_eff}
\end{equation}
and is rigidly rotated by $-\theta/2$ relative to the exact circle,
as shown in Fig.~\ref{fig:effective_circle}.
The Trotterized dynamics
is therefore equivalent to the exact dynamics of the QRM with two
geometric modifications: a rescaling of the coupling strength by
$1/\sinc(\theta/2)$, and a rigid phase-space rotation by $-\theta/2$,
both encoded in the single complex number $\varepsilon(\theta)$.

This constitutes the central result of this section: the first-order
Trotterization of the QRM in the degenerate limit does not generate a
generic, accumulating approximation error. Instead, it reproduces
exactly the dynamics of the same model with renormalized parameters,
independently of the simulation time, the initial state, or the number of steps.

This geometric picture suggests a natural question: can the Trotterization error be corrected altogether? Reducing the coupling to $g\,\mathrm{sinc}(\theta/2)$ matches the radius of the Trotterized and exact circles but leaves the $-\theta/2$ rotation of Eq.~(\ref{eq:ratio}) uncorrected, so the two trajectories still do not coincide. Compensating this rotation only requires adding the opposite phase to the coupling,
\begin{equation}
 g \;\to\; \tilde g = g\,\mathrm{sinc}(\theta/2)\,e^{i\theta/2},
 \label{eq:g_correction}
\end{equation}
equivalent to the interaction $\sigma_x(a+a^\dagger)\to\sigma_x(ae^{-i\theta/2}+a^\dagger e^{i\theta/2})$. With this choice, $\gamma_k=\alpha(t_k)$ exactly for every $k$: the Trotterized and exact trajectories coincide at all times, and the simulation is exact. Since the interaction is implemented as a synthesized gate rather than a fixed physical coupling, this complex amplitude is fully accessible through standard phase control of the pulse implementing each Trotter step, at no additional circuit cost.

\section{Fidelity and Trotterization error}
\label{sec:fidelity}

The phase-space picture developed in the previous section provides an intuitive visualization of the Trotterization error, but $\Delta_k$ as 
defined there is a property of the operator — not of the full quantum state. To obtain a complete characterization of the simulation error, 
we now compute the state fidelity
\begin{equation}
 \mathcal{F}(t_k) = \left|\langle\psi_\mathrm{ex}(t_k)|
 \psi_\mathrm{tr}(t_k)\rangle\right|^2,
 \label{eq:fidelity}
\end{equation}
which measures the global overlap between the exact and Trotterized states at each time step. As we show below, in the degenerate limit 
$\omega_0=0$ this quantity admits a closed-form expression entirely governed by $\Delta_k$, mirroring the structure of the survival probability $\mathcal{S}(t)$ derived in Sec.~\ref{sec:revivals}.

\subsection{Closed expression for the fidelity}

For an initial state $\ket{\psi_0}=c_+\ket{+,n}+c_-\ket{-,n}$, the exact state at $t=t_k$ follows from Eq.~(\ref{eq:Udag_tildeH_U_psi0}),
\begin{eqnarray}
 \ket{\psi_\mathrm{ex}(t_k)}
 &=&\sum_{s=\pm} c_s \ket{s}\, D\!\left[s \alpha(t_k)\right]\ket{n}.
\end{eqnarray}
The Trotterized evolution, using Eq.~(\ref{eq:Tk_exact}), reads
\begin{eqnarray}
\ket{\psi_\mathrm{tr}(t_k)} &=& \mathcal{T}^k \ket{\psi_0} = D(\gamma_k \sigma_x)e^{-ik\theta a^\dagger a}\ket{\psi_0}\nonumber\\
&=& e^{-ikn\theta}\sum_{s=\pm} c_s \ket{s} D(s\gamma_k) \ket{n},
\end{eqnarray}
where the global phase $e^{-ikn\theta}$ is irrelevant for the fidelity and will be omitted from here on.

\begin{figure}[t]
\centering
\includegraphics[width=\columnwidth]{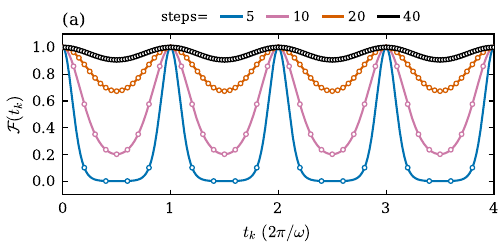}
\includegraphics[width=\columnwidth]{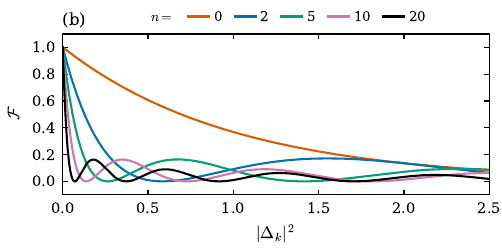}
\caption{Fidelity $\mathcal{F}$ as a function of simulation parameters for coupling ratio $\lambda=g/\omega=2$.
(a) Fidelity $\mathcal{F}(t_k)$ vs.\ time for an initial state with $n=0$ photons and different numbers of Trotter steps per period, showing exact
revivals at multiples of the oscillation period $T=2\pi/\omega$, independently of the step size. The fidelity decay between revivals decreases as
the number of steps increases, consistent with convergence to the exact dynamics in the continuum limit $\delta t\to 0$.
Open circles mark the actual (discrete) Trotterized values $\mathcal{F}(t_k)$;
for each curve, the solid line is the smooth function obtained by formally extending Eq.~(\ref{eq:fidelity_final}) to continuous $t$ at that curve's own step size $\theta=2\pi/n_\mathrm{steps}$ (since $n_\mathrm{steps}\,\delta t=T$), shown only to reveal the oscillation pattern between the discrete points, with no independent physical meaning at intermediate times.
(b) Fidelity $\mathcal{F} = e^{-|\Delta_k|^2}[L_n(|\Delta_k|^2)]^2$ as a function of $|\Delta_k|^2$, shown as a continuous curve to illustrate the
functional dependence for increasing initial photon number $n$. Higher Fock states exhibit enhanced sensitivity to
Trotterization errors through the Laguerre factor $L_n$, with additional zeros appearing for $n>0$ that reflect the non-classical
structure of displaced Fock states.}
\label{fig:fidelity}
\end{figure}

The overlap entering the fidelity can be evaluated straightforwardly.
The cross terms vanish by the orthogonality of the $\ket{\pm}$ states, and since the diagonal matrix element $\langle n|D(\xi)|n\rangle$ 
depends only on $|\xi|^2$, both terms $s=\pm$ contribute equally. Using $|c_+|^2+|c_-|^2=1$ and the composition rule of displacement 
operators, the overlap reduces, up to a phase factor, to a single bosonic matrix element,
\begin{eqnarray}
 \langle \psi_\mathrm{ex}(t_k) | \psi_\mathrm{tr}(t_k) \rangle
 &=& \bra{n} D^\dagger\![\alpha(t_k)]
     D(\gamma_k)\ket{n} \nonumber\\
 &=& \bra{n} D(\Delta_k)\ket{n} \nonumber\\
 &=& e^{-|\Delta_k|^2/2}\,L_n(|\Delta_k|^2),
 \label{eq:overlap_exact}
\end{eqnarray}
where $L_n(x)$ denotes the Laguerre polynomial of order $n$ and $\Delta_k$ is defined in Eq.~(\ref{eq:Deltak}). 
Taking the modulus squared yields 
\begin{equation}
\mathcal{F}(t_k) = e^{-|\Delta_k|^2}\left[L_n(|\Delta_k|^2)\right]^2.
\label{eq:fidelity_final}
\end{equation}
The fidelity is therefore fully controlled by the single real parameter $|\Delta_k|^2$, see Eq. (\ref{eq:Delta_factored}), which factorizes as 
\begin{equation}
|\Delta_k|^2 = |\varepsilon(\theta)-1|^2\,|\alpha(t_k)|^2,
\label{eq:Delta_k_sq}
\end{equation}
directly proportional to $|\alpha(t_k)|^2$ with a $k$-independent weight. Since $|\alpha(t_k)|^2$ is precisely the quantity controlling the physical revivals, Eq.~(\ref{eq:alpha_square}), it follows that $|\Delta_k|^2$ vanishes at exactly the same instants $t_k=2\pi\ell/\omega$, independently of the Trotter step size, while the depth of the fidelity decay between revivals is set by $|\varepsilon(\theta)-1|^2$, as shown in Fig.~\ref{fig:fidelity}.
For $n=0$, $L_0=1$ and the fidelity reduces to the purely geometric expression $\mathcal{F}_0=e^{-|\Delta_k|^2}$; for $n>0$, higher Fock states are more sensitive to Trotterization errors through the factor $[L_n(|\Delta_k|^2)]^2\leq1$.

\subsection{Structural identity with the physical revivals}

Written as matrix elements, $\mathcal{S}(t)$ and $\mathcal{F}(t_k)$ are the same mathematical object: the overlap of a Fock state with a displaced copy of itself,
\begin{equation}
 \mathcal{S}(t) = \big|\langle n|D[\alpha(t)]|n\rangle\big|^2, \qquad
 \mathcal{F}(t_k) = \big|\langle n|D(\Delta_k)|n\rangle\big|^2, \nonumber
\end{equation}
differing only in which displacement drives them. For $\mathcal{S}(t)$, the displacement is $\alpha(t)$, the exact phase-space amplitude measured from the origin — how far the physical system has moved from its initial state. For $\mathcal{F}(t_k)$, the displacement is $\Delta_k$, the phase-space separation between the exact and Trotterized amplitudes — how far the simulation has drifted from the system it is meant to reproduce. This identification is what makes both quantities collapse to the same closed form, $e^{-|\xi|^2}[L_n(|\xi|^2)]^2$, Eqs.~(\ref{eq:survival_Ln}) and (\ref{eq:fidelity_final}): a physical revival and a fidelity revival are the same question — the distinguishability of a displaced Fock state — asked with two different displacements.

This equivalence relies on two independent reductions established earlier in this work: $\mathcal{S}(t)$ takes this form only because $H$ is exactly diagonalized (Sec.~\ref{sec:diagonalization}), while $\mathcal{F}(t_k)$ does so only because $\mathcal{T}^k$ admits the closed form (Sec.~\ref{sec:trotter}). Neither is guaranteed a priori, and each is a genuine
result of this work.

\section{Conclusions and Outlook}

We have presented an exact analytical characterization of first-order Trotterization of the quantum Rabi model in the degenerate limit $\omega_0=0$. The Trotter operator admits a closed-form expression, $\mathcal{T}^k=D(\gamma_k\sigma_x)e^{-ik\theta a^\dagger a}$, from which the phase-space error is shown to reduce to a single complex number, $\Delta_k=[\varepsilon(\theta)-1]\alpha(t_k)$, proportional to the exact trajectory itself. This proportionality reveals that the Trotterized dynamics does not accumulate a generic error, but reproduces exactly the dynamics of the same model with a renormalized coupling and a fixed phase-space rotation. The resulting state fidelity $\mathcal{F}(t_k)$ admits an equally closed form, structurally identical to the physical survival probability $\mathcal{S}(t)$: both are diagonal matrix elements of a displacement operator, $|\langle n|D(\xi)|n\rangle|^2$, evaluated at two different phase-space distances. Physical revivals and fidelity revivals are, in this sense, the same geometric question asked of two different pairs of points.

This analysis is restricted throughout to the exactly solvable degenerate limit $\omega_0=0$, which isolates the essential geometric mechanism at the cost of leaving the experimentally relevant regime $\omega_0\neq0$ outside its scope. The extension to $\omega_0\neq0$, where the closed-form results derived here are expected to hold only approximately, is a natural direction for future work, to be addressed numerically.

A further direction concerns the exact correction identified in Sec.~\ref{sec:effective}: promoting the Trotter-step coupling to a complex amplitude, $g\to g\,\mathrm{sinc}(\theta/2)\,e^{i\theta/2}$, exactly cancels the geometric error at no additional circuit cost. Since this amplitude is a synthesized gate parameter rather than a fixed physical coupling, this correction is in principle accessible on current digital quantum hardware, and its experimental validation — together with a possible extension of the same strategy beyond the degenerate limit — is left for future work.

\begin{acknowledgments}
This research was funded by the Basque Government through Grant No. IT1887-26
and
Spanish Government MCIU through Grant No. PID2021-126273NB-I00.
\end{acknowledgments}

\appendix
\section{Survival probability}
\label{app:survival}

In this appendix we derive an explicit expression for the probability 
amplitude 
entering 
Eq.~(\ref{eq:survival_prob_general}), for a general initial state 
$\ket{\psi_0}=c_+\ket{+,n}+c_-\ket{-,n}$.

Using the transformation $\mathcal{U}=D(\lambda\sigma_x)$ that diagonalizes 
the Hamiltonian, $\widetilde{H}=\mathcal{U} H\mathcal{U}^\dagger $, the 
amplitude can be written as
\begin{equation}
 \mathcal{A}(t) \equiv \langle\psi_0|e^{-iHt}|\psi_0\rangle
 = \bra{\psi_0}\mathcal{U}^\dagger e^{-i\widetilde{H}t}\mathcal{U}\ket{\psi_0}.
\end{equation}

We evaluate this expression in four steps:

\begin{itemize}
 \item [(i)] Action of $\mathcal{U}$.  
 Using $\sigma_x\ket{\pm}=\pm\ket{\pm}$,
\begin{equation}
 \mathcal{U}\ket{\psi_0} = \sum_{s=\pm} c_s \ket{s}\, D(s\lambda)\ket{n}.
\end{equation}

\item [(ii)] Time evolution in the diagonal frame. 
Since $\widetilde{H}$ is diagonal, the evolution acts trivially on the displaced Fock states, 
amounting to a phase-space rotation of the displacement amplitude,
\begin{equation}
 e^{-i\widetilde{H}t}\,\mathcal{U}\ket{\psi_0} 
 = \sum_{s=\pm} c_s \ket{s}\, D(s\lambda\, e^{-i\omega t})\ket{n}.
 \label{eq:psi_tilde_t}
\end{equation}

\item [(iii)] Return to the original frame.
Applying $\mathcal{U}^\dagger=D(-\lambda\sigma_x)$ and using the composition rule of 
displacement operators,
\begin{equation}
 \mathcal{U}^\dagger e^{-i\widetilde{H}t}\,\mathcal{U}\ket{\psi_0}
 = \sum_{s=\pm} c_s \ket{s}\, D\!\left[s\lambda(e^{-i\omega t}-1)\right]\ket{n},
 \label{eq:Udag_tildeH_U_psi0}
\end{equation}
up to an irrelevant global phase. 

\item [(iv)] Projecting onto $\bra{\psi_0}$ and using the orthogonality of the $\ket{\pm}$ states,
\begin{equation}
 \mathcal{A}(t) = \sum_{s=\pm} |c_s|^2\,
 \bra{n}D\!\left(s\lambda(e^{-i\omega t}-1)\right)\ket{n}.
\end{equation}
Since the diagonal matrix element $\langle n|D(\xi)|n\rangle$ depends only on $|\xi|^2$, both terms 
$s=\pm$ are equal, and using $|c_+|^2+|c_-|^2=1$ the sum collapses to a single bosonic matrix element,
\begin{equation}
 \mathcal{A}(t) = \bra{n}D\!\left(\alpha\right)\ket{n},
 \qquad \alpha \equiv \lambda(e^{-i\omega t}-1),
\end{equation}
independent of $c_+$ and $c_-$. This is the probability amplitude used in 
the main text.

\end{itemize}


\end{document}